# Ultrafast Reconfigurable Topological Photonic Processing Accelerator

**Wenfeng Zhou**[1,4], **Xin Wang**[1,4], **Xun Zhang**[1,4], **Yuqi Chen**[1], **Min Sun**[1], **Jingchi Li**[1], **Xiong Ni**[1], **Yahui Zhu**[1], **Qingqing Han**[1], **Jungan Wang**[2], **Chen Yang**[2], **Bin Li**[2], **Feng Qiu**[2,3], **Yikai Su**[1*], **Yong Zhang**[1*]

1. State Key Laboratory of Photonics and Communications, Department of Electronic Engineering, Shanghai Jiao Tong University; Shanghai, China.
2. AIRC, Hangzhou Institute for Advanced Study, University of Chinese Academy of Sciences; Hangzhou, China.
3. Juhe Electro-optic (Hangzhou) Tech. Co. Ltd., Hangzhou, China.
4. These authors contributed equally.

*Corresponding author: Yikai Su and Yong Zhang (email: yikaisu@sjtu.edu.cn; yongzhang@sjtu.edu.cn)

## Abstract

The rise of artificial intelligence has triggered exponential growth in data volume, demanding rapid and efficient processing. High-speed, energy-efficient, and parallel-scalable computing hardware is thus increasingly critical. We demonstrate a wafer-scale non-volatile topological photonic computing chip using topological modulators. Leveraging the GHz-speed electro-optic response and nonvolatility of ferroelectric lead zirconate titanate (PZT) thin films via topological photonic confinement, our chip enables 1,000× accelerated reconfiguration, zero-static-power operation, and a computational density of 266 trillion operations per second per square millimeter (TOPS/mm²). This density surpasses that of silicon photonic reconfigurable computing chips by two orders of magnitude and thin-film lithium niobate platforms by four orders of magnitude. A 16-channel wavelength-space multiplexed chip delivers 1.92 TOPS throughput with 95.64% digit-recognition accuracy and 94.5% precision for solving time-varying partial differential equations. Additionally, the chip supports functional reconfiguration for high bandwidth density optical I/O. This work establishes ferroelectric topological photonics for efficient high-speed photonic tensor processing.

Explosive artificial intelligence (AI) growth in autonomous driving, industrial internet of things, and medical diagnostics demands computing architectures delivering simultaneous real-time processing, ultra-high efficiency, and hyperscale capability[1-4]. Von Neumann-based digital processors face fundamental barriers: interconnect RC delays and Joule heating cause superlinear power growth with frequency, creating insurmountable performance scaling limits[5,6].

Photonic computing emerges as a transformative computing paradigm with three fundamental advantages: terahertz-level operating bandwidth, inherent parallelism that leverages the frequency and polarization degrees of freedom, and low latency[7-10]. In recent years, photonic computing has achieved remarkable breakthroughs, demonstrating revolutionary advantages in computational speed, energy efficiency, and latency performance[11-15].

Despite recent advances, photonic computing architectures face three fundamental challenges: 1) Computation capacity and density. While on-chip diffractive optical neural networks achieve >100 trillion operations per second (TOPS) throughput with high compute density[16], their static weights inherently lack reconfigurability. Reconfigurable non-diffractive networks (e.g., Mach–Zehnder interferometer (MZI) meshes or microring banks) scale capacity via parallelization[17], yet remain constrained by physical channel expansion rather than device innovation. This approach inevitably enlarges footprints[5], capping compute density at <2 TOPS/mm².

2) Reconfigurability. Constrained by microsecond-scale thermo-optic weight tuning latency[18-20], most silicon/silicon nitride architectures lack rapid parameter updating capability. This slow reconfiguration prevents dynamic input/output dimension adaptation[21], critically impeding real-time fan-in/fan-out reconfiguration for computationally intensive tasks such as time-dependent partial differential equations (PDEs) solving[22-24].

3) Energy efficiency. Multi-processor AI systems demand ultra-efficient hardware. Milliwatt-level power consumption in thermo-optic heaters and carrier-based modulation contradicts photonic computing's energy-efficiency advantage, severely limiting scalable integration and practical deployment[25-27].

Emerging material platforms have recently been explored to address these challenges. While thin-film lithium niobate (TFLN) electro-optic modulators enable nanosecond reconfiguration[28,29], their low Pockels coefficient necessitates centimeter-scale modulator arms[30], resulting in computational densities two orders of magnitude lower than silicon photonic devices[31]. DC drift further necessitates power-hungry thermal phase shifters. Phase-change materials provide nonvolatility and zero static power[5], yet suffer from slow state transitions and optical loss during amorphous-crystalline transitions[32]. Precise multilevel control remains challenging[33]. Consequently, developing photonic computing hardware that achieves high computation density, dynamic configurability, and energy efficiency remains a challenge.

Here we demonstrate a wafer-scale non-volatile topological photonic computing (NTPC) chip monolithically integrating topological modulators on a 4-inch thin-film PZT platform. Leveraging the 67-GHz-speed electro-optic response and nonvolatility of ferroelectric PZT thin film, our chip achieves optical path control with 1,000× faster dynamic reconfiguration and zero static power consumption. By integrating 20 ultra-compact topological modulators via 16-channel wavelength-space multiplexing, the NTPC delivers 1.92 TOPS throughput at 266 TOPS/mm² computational density, outperforming silicon reconfigurable computing chips by two orders of magnitude and TFLN chips by four orders of magnitude. Successful applications include image edge detection, handwritten digit recognition (95.64% accuracy), and 2D heat transfer solutions (94.5% accuracy), establishing the first high-speed and non-volatile ferroelectric platform for photonic tensor processing. Furthermore, non-volatility enables dynamic reconfiguration of the NTPC. This allows the monolithically integrated topological modulators to perform dense wavelength division multiplexing (DWDM), achieving optical I/O with a shoreline bandwidth density exceeding 3.56 Tbps/mm. These achievements underscore the versatile adaptability and functional extensibility of the NTPC.

## Results

**Accelerator architecture and operating principle.** We present an NTPC that synergistically integrates wavelength- and space-division multiplexing (WDM and SDM). The NTPC enables 16-channel parallel processing by combining 4 wavelength channels with 4 spatial modes (Fig. 1d), forming a dense multiply-accumulate array. Each computing unit comprises two serially coupled topological photonic crystal (TPC) modulators. The front-end modulator performs dynamic encoding of input data, while the back-end modulator enables real-time weight updates.

The NTPC's core component is high-bandwidth electro-optic modulators. On-chip electro-optic modulators typically use silicon's carrier dispersion or TFLN's Pockels effect. Silicon modulators face limited bandwidth, nonlinearity, and carrier absorption losses. TFLN modulators, despite their potential, exhibit low electro-optic coefficients (~30 pm/V), requiring centimeter-scale arms that compromise compactness and computational density (Fig. 1a). To overcome these limitations, we employ thin-film PZT with a significantly higher electro-optic coefficient (~100 pm/V) to achieve highly efficient electro-optic modulation[34]. Furthermore, we utilize topological photonic crystal microcavities featuring smaller mode volumes (see Supplementary Section [1] for details). This design enhances optical field confinement, strengthens the electro-optic interaction, and thus enables more efficient modulation. Consequently, the required modulation length is reduced, leading to higher computational density.

The NTPC requires the cascading of two electro-optic modulators, necessitating precise wavelength alignment between them. Wavelength alignment is conventionally performed via the thermo-optic effect (Fig. 1a). Unfortunately, this approach incurs significant power consumption, often in the tens of milliwatts per unit[35]. As chip scale increases, this power demand escalates dramatically and introduces detrimental thermal crosstalk[36]. We overcome these limitations by exploiting the intrinsic non-volatility of thin-film PZT to enable near-zero-power tuning and reconfiguration (see Supplementary Section [2] for non-volatile operation principle). PZT offers a solution through direct electrical manipulation of ferroelectric domain polarization[37], enabling precise refractive index control without requiring sustained bias voltages (Fig. 1b).

To demonstrate the NTPC's performance benefits, we evaluate three key applications (Fig. 1c): image processing, handwritten digit classification, and solving time-dependent PDEs. The NTPC achieves orders-of-magnitude improvements in both computational density and fan-in size (Fig. 1e). Fabrication, characterization of the NTPC, along with details of the photonic computing demonstrations, are presented in subsequent sections.

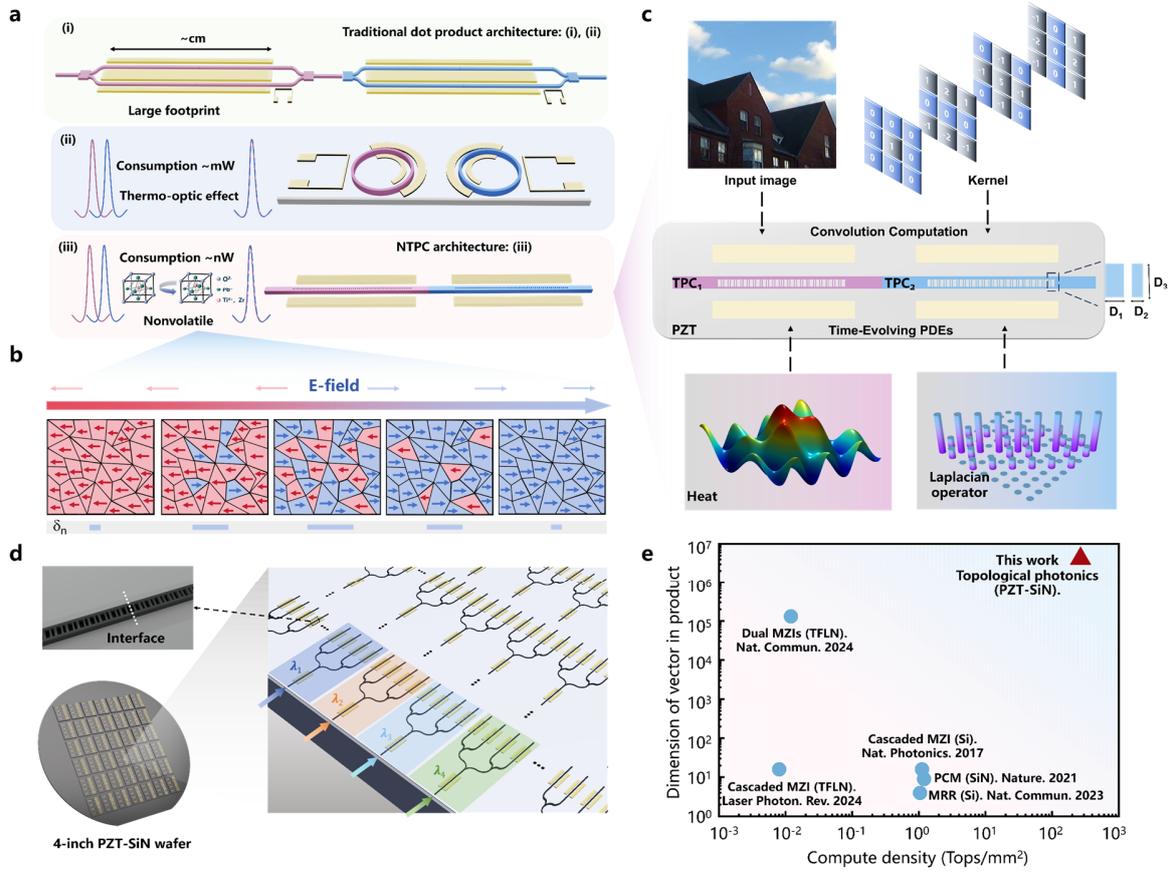

**Fig. 1. Concept of integrated non-volatile topological photonic computing chip (NTPC).** (**a**) Traditional dot product architecture (i), (ii), and NTPC architecture (iii). (**b**) Schematic of ferroelectric domains in PZT (top view) and the relationship between effective refractive index $\delta_n$ and polarization electric field strength. (**c**) The computing unit performs two independent operations: 1) optical convolution with images encoded on $TPC_1$ and kernels on $TPC_2$, and 2) heat conduction simulation via heat source loaded on $TPC_1$ and the Laplace operator configured on $TPC_2$. (**d**) Schematic of the NTPC chip with 16-channel parallel processing. (**e**) Performance comparison of the dimension of vector in product and compute density among state-of-the-art platforms and architectures.

**Fabrication and characterization of NTPC.** This section details the fabrication and characterization of the NTPC chip. Crack-free PZT thin films with preferential (100) orientation are initially deposited on 4-inch $SiO_2$/Si substrates using seed layers, employing a solution-based chemical deposition technique[38]. This method provides excellent compatibility with complementary metal-oxide-semiconductor (CMOS) technology while offering advantages in scalability and cost-effective manufacturing (Fig. 2a). To overcome challenges in etching thin-film PZT—specifically achieving vertical sidewalls—we implement silicon nitride (SiN) loaded PZT waveguides (Fig. 2c). Leveraging the well-established fabrication processes for SiN offers a viable pathway for the large-scale integration of thin-film PZT photonic devices (see methods for details).

To fully exploit the massive parallelism inherent in photonic computing, we implement a 16-channel design utilizing 4-channel WDM and 4-channel SDM (Fig. 2b). Four front-end topological electro-optic modulators dynamically encode input data for the four distinct wavelengths. Sixteen back-end topological modulators perform real-time weight updates across the four wavelengths and four spatial dimensions. Both modulator types utilize the high-speed Pockels effect in thin-film PZT, enabling rapid signal encoding and real-time reconfiguration. The fabricated topological cavity exhibits a Q factor of 9,000 and an extinction ratio of 24 dB (Fig. 2d), while the fabricated 3-dB splitter shows an excess loss of 0.5 dB (results of unit devices are detailed in the Supplementary Section [3]). Our 4-inch PZT wafer fabrication technology supports further scalability through additional multiplexing dimensions, such as polarization and mode, promising significantly enhanced on-chip computational capacity.

**High-speed topological modulators of NTPC.** High-speed electro-optic modulators are the core functional elements of the NTPC. To concurrently maximize modulation bandwidth and minimize footprint, we deploy a dual-pronged approach: 1) Material Optimization: Leveraging PZT's superior electro-optic coefficients enables highly efficient modulation, significantly reducing the required optical interaction length. 2) Structural Innovation: Employing topological photonic crystal microcavities achieves exceptional optical field confinement, minimizing mode volume. This dramatically intensifies the electro-optic interaction, facilitating compact, high-speed modulation.

The modulator operates via a topological interface state formed at the junction of two one-dimensional (1D) TPCs with distinct topological invariants, engineered using the Su-Schrieffer-Heeger (SSH) model[39]. Unlike conventional photonic crystal nanobeam cavities prone to multiple resonant modes, our topological cavity design offers independent control over the Q factor and mode volume while rigorously maintaining intrinsic single-mode operation, thereby eliminating complex mode management. The NTPC requires multiple wavelength-specific topological modulators, achieved by adjusting lattice periods. Figure 2e presents measured transmission spectra for four TPCs with varying periods, each revealing a sharp resonant peak corresponding to a topological interface state at a unique wavelength. Data fitting yields a Q-factor of 9000, corresponding to a photon lifetime of $\tau = Q\lambda/(2\pi c) \approx 7.3$ ps, which theoretically sets an upper modulation bandwidth limit of ~22 GHz.

To activate the electro-optic effect in the TPC modulator, a square-wave pulse train with a period of 1 s and a 50% duty cycle is applied to the electrodes. This poling process continues for 15 minutes to ensure full and uniform alignment of the ferroelectric domains along a single preferred direction. The static tuning efficiency is then characterized by sweeping the DC bias voltage while monitoring shifts in the transmission spectrum, yielding a value of 14 pm/V (see Supplementary Section [4] for details). Exploiting this topological interface state enables the realization of the first topological PZT modulator, achieving an ultra-compact footprint of 1.6 × 225 μm². Harnessing transient peak response enables the modulator to surpass the photon-lifetime bandwidth limit, reaching over 67 GHz. To our knowledge, this is the most compact thin-film PZT modulator demonstrated with over 40 GHz bandwidth. All four wavelength-specific modulators consistently achieve modulation bandwidths >67 GHz (Figs. 2f-i), demonstrating exceptional stability and reproducibility (see Supplementary Section [5] for details).

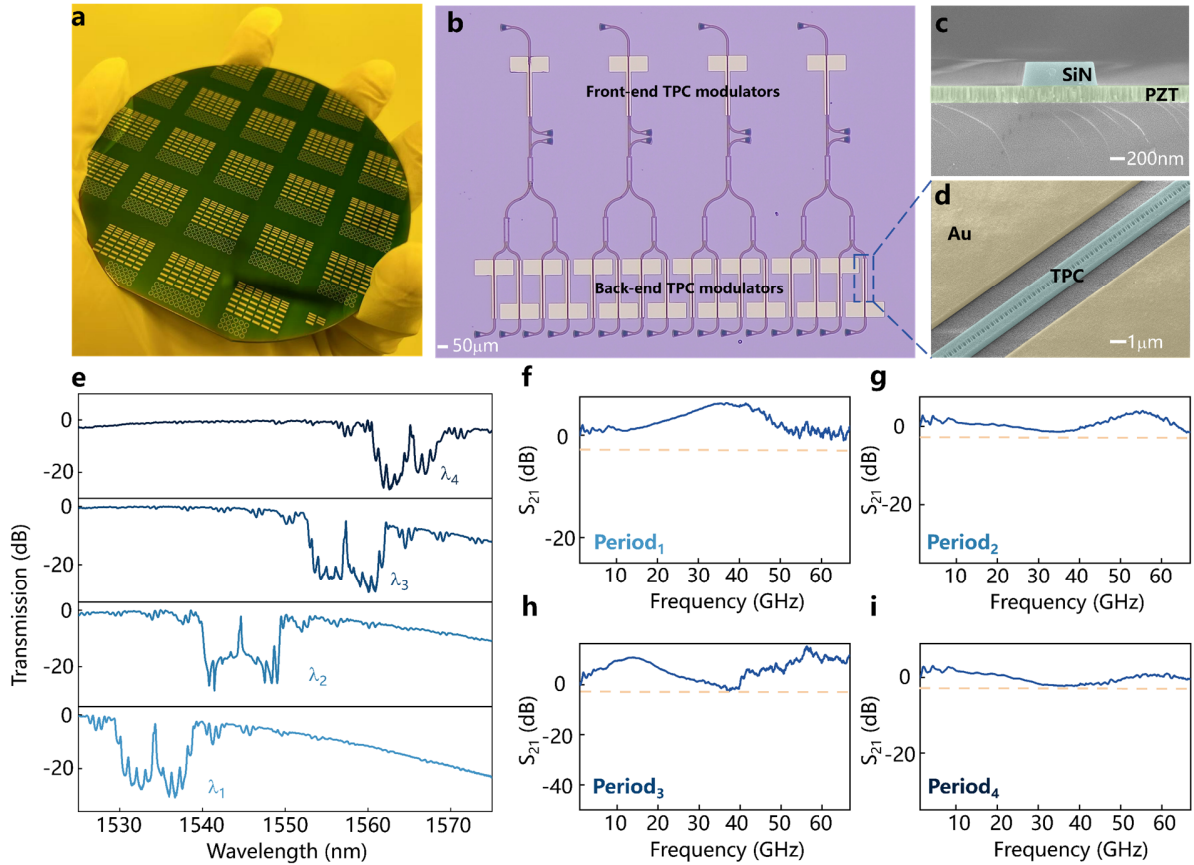

**Fig. 2. Characterization of the NTPC.** (**a**) A 4-inch SiN-loaded PZT wafer containing NTPC. (**b**) Microscopy image of the NTPC chip. False-color scanning electron microscope (SEM) images of (**c**) SiN-loaded PZT waveguide cross-section and (**d**) topological modulator. (**e**) Transmission spectra of the TPCs across different periods, with resonances at 1534 nm, 1545 nm, 1555 nm, and 1566 nm. (**f~i**) Measured $S_{21}$ responses for topological modulators with different periods.

**Non-volatile reconfiguration operation of NTPC.** In the 16-channel NTPC chip, each channel requires two electro-optic modulators at the same wavelength: a front-end TPC modulator for high-speed signal encoding and a back-end modulator for real-time weight updates. Thermo-optic tuning is a common method to achieve wavelength alignment, but integrated microheaters consume over 10 mW per device, raising photonic computing's energy costs. To address this challenge, we employ the non-volatile characteristics of PZT to achieve wavelength calibration.

We perform non-volatile testing via a $V_{set}$ sweep (9 V to 32 V, Fig. 3b), achieving 23 tunable non-volatile states

(Fig. 3c). This continuously tunable refractive index enables us to achieve arbitrary wavelength adjustments across a 3 nm range. As a typical demonstration of non-volatility, we showcase a stable six-level memristor. Defining 550 pm detuning as one state enables 6 programmable operations (Fig. 3d). To verify repeatability (critical for NTPC performance), we conduct 10 erase-write cycles on these states, observing <50 pm wavelength variation throughout (Fig. 3e), confirming excellent stability.

We add two monitor ports (#M1 and #M2) to track alignment between $TPC_1$ and $TPC_2$ microcavities. Fifteen percent of the light is routed to monitoring ports via a custom-designed beam splitter, enabling real-time resonance tracking (Fig. 3a). Initial measurements show a 1700 pm offset between $TPC_1$ (1544.1 nm) and $TPC_2$ (1545.8 nm) due to fabrication variations (Fig. 3f). This misalignment creates a combined bandgap that blocks Path3 transmission, disrupting the multiply-accumulate operation between input data and weights. The spectral alignment protocol employs controlled red-shifting of $TPC_1$'s resonance through applied bias voltages. The applied voltage (60 s duration) induces ferroelectric domain switching, followed by a 60 s stabilization period to ensure domain relaxation before Path3 spectral characterization. Based on the non-volatile characteristics of PZT, we estimate that the tuning voltage should fall within the 23-25 V range. Systematically:(i) Without bias (0 V), Path3 exhibits no resonant transmission. (ii) At 23 V bias, partial spectral overlap occurs between $TPC_1$ and $TPC_2$, yielding resonance peaks with 12 dB extinction ratio. (iii) At 25 V, $TPC_1$ undergoes a 1.7 nm red-shift, achieving complete spectral alignment with $TPC_2$ and significantly enhanced resonance (22-dB extinction ratio; Fig. 3g). After each measurement cycle, a reverse bias resets the device to its initial state. This voltage-dependent spectral tuning demonstrates precise control over cavity-cavity coupling in the photonic network. PZT-based non-volatile alignment ensures critical long-term stability for optical chips. Our 25-hour monitoring of Path3 shows sustained 22 dB extinction ratios and <50 pm wavelength drift (Fig. 3h), confirming the method's reliability.

Compared to conventional thermo-optic tuning—which typically consumes tens of milliwatts per device—our PZT-based ferroelectric non-volatile alignment requires merely 0.05 nW total energy, with a tuning efficiency of 0.0294 nW/nm (see Supplementary Section [6] for detailed analysis). This represents an eight-order-of-magnitude reduction in power consumption. The technology thereby provides a robust solution for optical computing chips, simultaneously delivering ultra-low power consumption and high computational density.

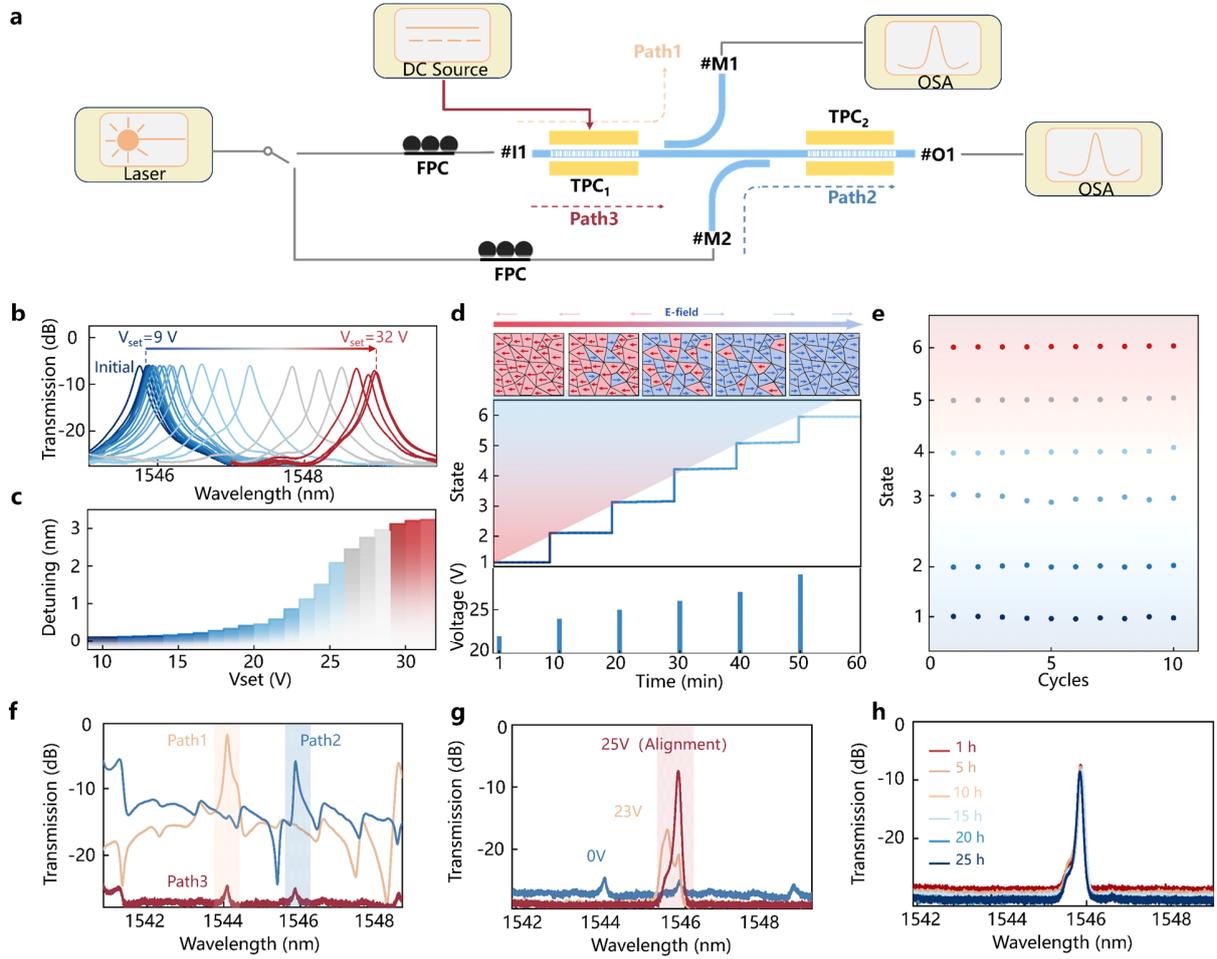

**Fig. 3. Non-volatile optical path reconfiguration of NTPC.** (**a**) Schematic of the measurement procedure. Path1: When light is injected into port #I1, the transmission spectrum of $TPC_1$ is detected at monitor port #M1; Path2: When light enters port #O1, the transmission spectrum of $TPC_2$ is observed at monitor port #M2; Path3: When light is injected at port #I1, the composite transmission spectrum resulting from sequential propagation through $TPC_1$ and $TPC_2$ is measured at output port #O1. FPC: fiber polarization controller; OSA: optical spectrum analyzer; DC Source: direct current source. (**b**) Transmission spectra under different applied voltages. (**c**) Statistical analysis of wavelength detuning under 23 distinct voltage conditions. (**d**) Top: Six non-volatile distinct states. Bottom: Applied voltage amplitudes. (**e**) Six distinct non-volatile states stability test through 10 erase-write cycles. (**f**) Transmission spectra of Path1, Path2, and Path3 under initial random fabrication variations. (**g**) Non-volatile transmission spectra of Path3 versus applied voltage. (**h**) Stability characterization of Path3 transmission spectra over 25 h after non-volatile alignment.

**NTPC for image processing tasks and handwritten digit recognition.** Our proposed programmable topological chip serves as a versatile platform for diverse optical functionalities. To rigorously assess its performance in convolutional computing, we carry out comprehensive system-level validation through image edge detection and a ten-class handwritten digit classification task, demonstrating its broad applicability and scalability.

Employing four distinct wavelengths combined with four-channel space-division multiplexing, we construct 16 parallel computing channels. During the preprocessing stage, a raw 512×512-pixel image is flattened into a 1×262144-dimensional feature vector and is loaded into the front-end TPC modulators, while the parameters of a 3×3 convolution kernel are transformed into a weight vector of identical dimension and are loaded into the back-end TPC modulators at a rate of 60 GBaud (Fig. 4a). Detailed testing procedures are provided in the Supplementary Section [7].

The computational results demonstrate successful implementation of three fundamental image processing operations through convolutional kernel reconstruction: identity transformation, sharpening enhancement, and edge extraction (Fig. 4b). Using separable convolution methods, we independently compute image gradient features along X/Y directions, with subsequent feature fusion clearly revealing edge structural information. Experimental data show excellent agreement with theoretical predictions. Detailed edge detection methodology is provided in the Supplementary Section [8].

Furthermore, we develop an optical neural network-based handwritten digit classification system (Fig. 4c). The implementation process comprises: During input processing, 28×28-pixel images are flattened into 1×784 (28×28) vectors. These vectors undergo temporal intensity encoding on front-end modulators at 60 GBaud. Calculations show single-image processing requires 13.08 ns (784÷60 GBaud), yielding a theoretical throughput of 76.5 million images/second. The feature extraction stage employs 3×3 optical convolution kernels operating at 60 GBaud. After ReLU nonlinear activation, outputs are converted into 1×169 feature vectors, which subsequently pass through a 169×128 fully-connected layer in the electrical domain to produce 1×10 classification vectors (maximum index indicates recognition result). Notably, the dimensionality of each vector is set to 262,144, constrained by the specifications of our high-speed arbitrary waveform generator (AWG). The NTPC achieves an unprecedented 60 GHz weight update speed at vector product dimensions of 262144×16=4.2×10$^6$, representing a six-orders-of-magnitude enhancement over conventional optical computing approaches. Convolution principles are detailed in the Supplementary Section [9].

Experimental results show excellent agreement between measured (gray) and ideal (blue) convolution outputs in the 50-75 ns range (Fig. 4d), confirming NTPC's feature extraction capability. On MNIST datasets (60,000 training/10,000 test images), hardware training curves match software simulations (Fig. 4e), showing consistent exponential decay in cross-entropy loss. Final test accuracy reaches 95.64% (Fig. 4f), approaching the theoretical 96.46% (Fig. 4g), with the 0.82% difference attributable to system noise, modulator drift, detector nonlinearity, and EDFA noise.

The NTPC integrates 20 topological modulators using 4-channel WDM and 4-channel SDM, achieving a 16-parallel-channel optical computing architecture. The chip achieves a computing speed of 120 GOPS with a peak computational capacity of 1.92 TOPS (derived from 60×2×4×4 operations). On a compact active area of 0.0072 mm², it delivers an exceptional compute density of 266 TOPS/mm². By eliminating thermo-optic tuning power consumption and leveraging the intrinsically low-energy nature of electro-optic modulation and non-volatile reconfigurability, the NTPC achieves an exceptional energy efficiency of 265 fJ/OP (see Supplementary Section [10] for detailed analysis). This breakthrough paves the way for next-generation optical computing technologies capable of simultaneously delivering high-speed operation, record compute density, and ultra-low power consumption.

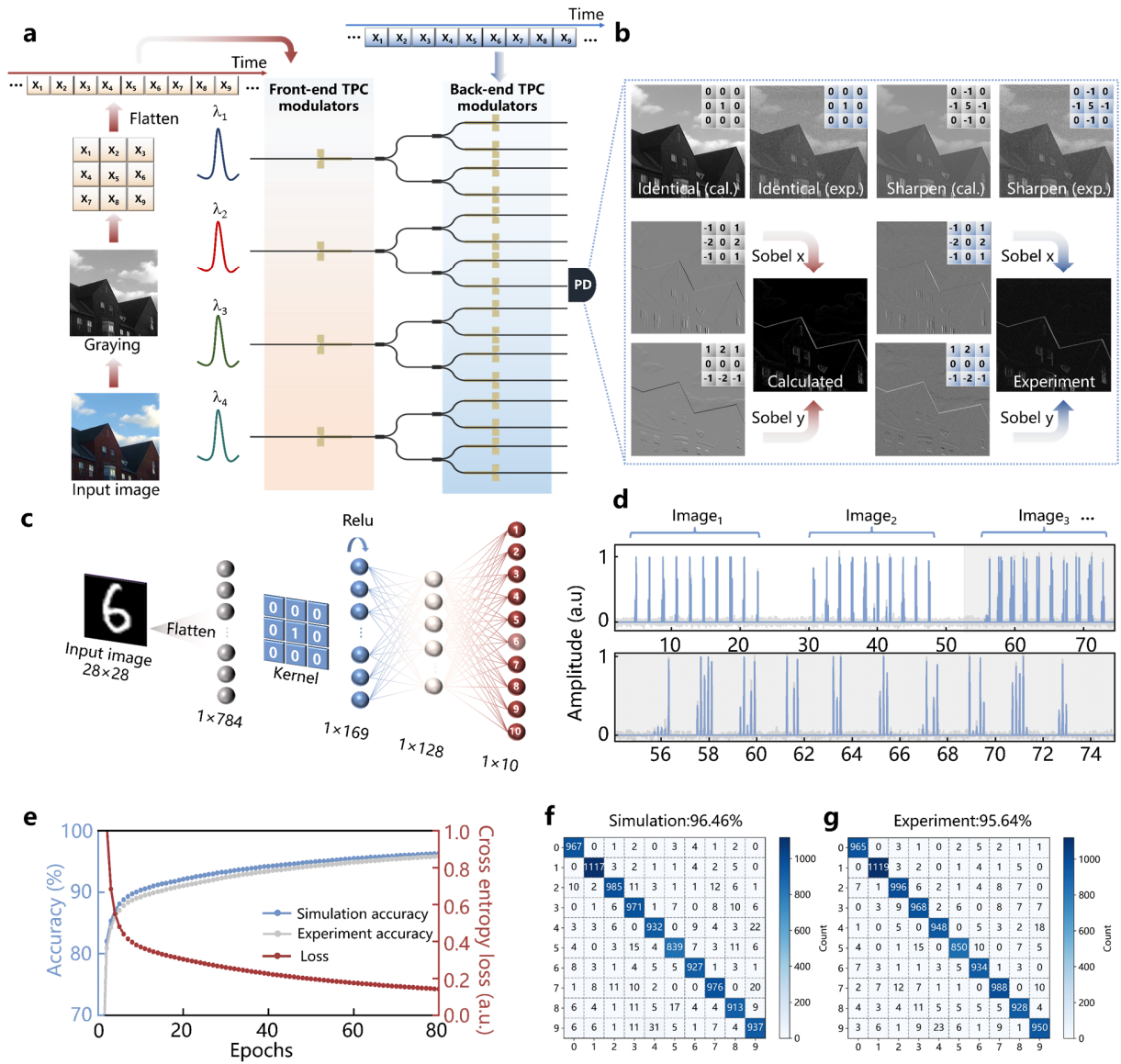

**Fig. 4. NTPC for image processing and digit classification tasks.** (**a**) Schematic diagram of NTPC for image processing tasks. (**b**) Convolution using 3×3 kernel: simulation and experimental results for identical image, sharpening, and edge detection. (**c**) Convolutional neural network framework for the handwritten digit identification system. (**d**) Ideal (blue) and experimental (gray) output waveforms of a convolutional operation performed on MNIST digit images using a 3×3 kernel. The magnified temporal profile captures the transient response from 55 ns to 75 ns. (**e**) Training dynamics over 80 epochs: comparative evolution of simulated accuracy, experimental accuracy, and experimental cross-entropy loss. (**f**) Calculated and (**g**) experimental confusion matrices (96.46% vs 95.64% accuracy).

**NTPC for solving partial differential equations.** Optical computing for time-evolving PDEs typically uses finite difference discretization, converting derivatives to matrix operations. The core computation involves iterative matrix-vector multiplications, with complexity scaling quadratically with grid size ($n^2$ variables → $n^2 \times n^2$ coefficient matrices). This creates scalability challenges for conventional hardware. The NTPC effectively resolves the challenge of exponential growth in device count resulting from high-resolution discretization through dynamic scaling of input/output matrix dimensions. This capability, therefore, positions NTPC as a highly promising platform for solving time-evolving PDEs.

The NTPC overcomes these limitations through flexible matrix sizing. It reshapes thermal field and coefficient matrices into 1D vectors: thermal data loads via front modulators, while coefficients load via rear modulators. Optical-domain multiplication results are captured and processed digitally to iteratively update solutions (Fig. 5a). Successful 6×6 grid Laplace operator demonstrations (Fig. 5b) highlight NTPC's advantages for optical PDE solving.

Data loading is configured at 1-GSa/s, with every 64 samples corresponding to one spatial point calculation. Figure 5c presents a comparative analysis of 1D thermal field evolution vectors between simulation (gray curves) and experimental computation (blue curves), where terminal fiducial markers facilitate data identification. Reconstructed 2D thermal field distributions derived from these 1D vectors are shown for both simulation (Fig. 5d) and experimental (Fig. 5e) results. Experimental data demonstrate good agreement with simulations across both dimensional representations.

Furthermore, the heat source is offset from the domain center to rigorously test distal heat transfer accuracy. Experimental measurements show good agreement with simulations, as evidenced by thermal field comparisons at 0.75 s, 2 s, and 3.5 s (Figs. 5f and 5g). The temperature evolution profile at grid point (5,5) (Fig. 5h) shows good agreement between simulation (light blue solid line) and experiment (red dashed line), validating our computational approach.

The evolution of computational accuracy exhibits several key characteristics. Initial precision is impacted by inherent instrument limitations, introducing random noise during measurements. As computation progressed, cumulative errors across successive time steps become increasingly pronounced, resulting in a gradual decline in overall accuracy (Fig. 5i). Complete technical details regarding data acquisition, processing pipeline, and precision quantification are provided in the Supplementary Section [11].

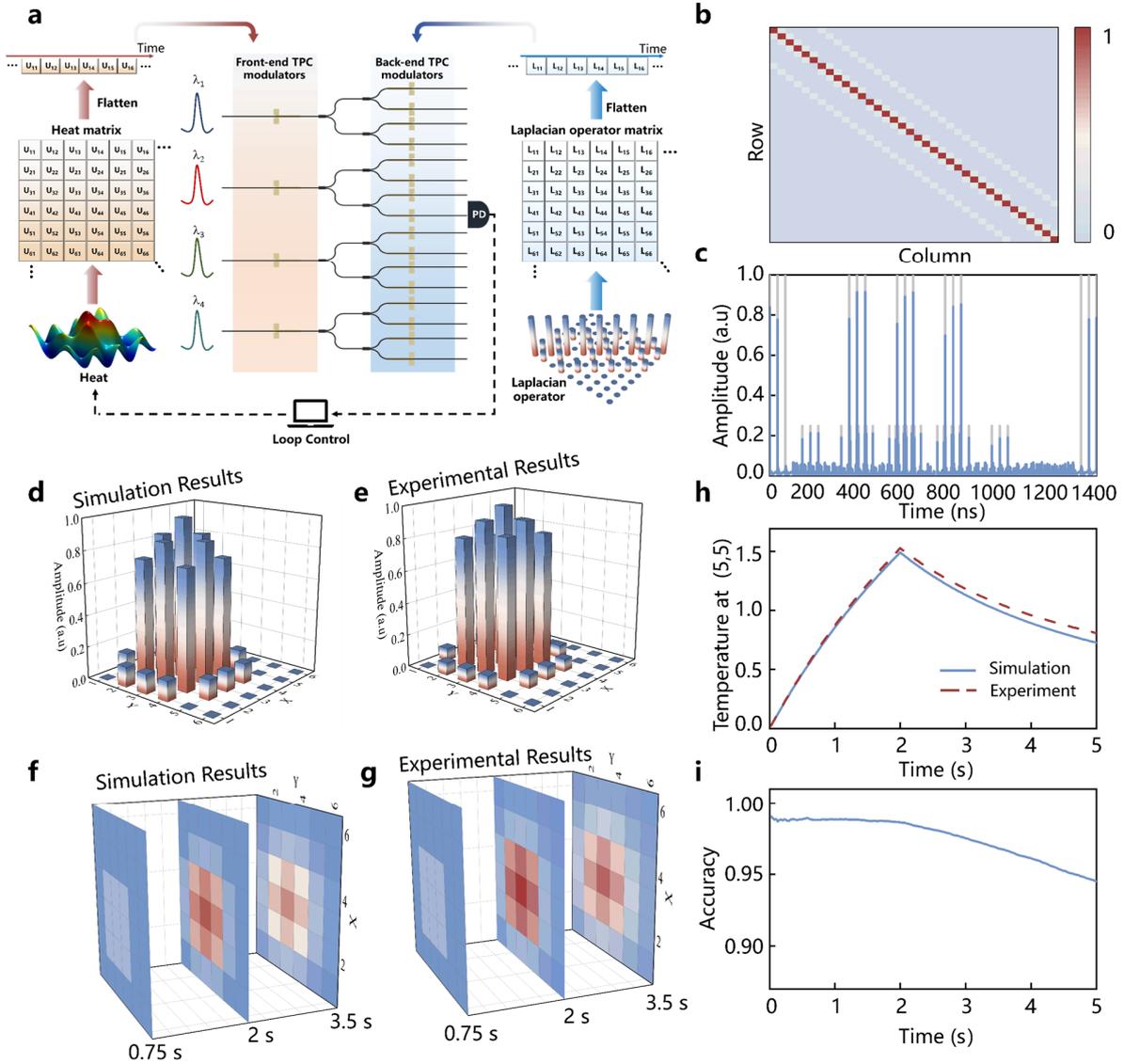

**Fig. 5. NTPC for Solving time-evolving partial differential equations - Heat Equation Solution.** (**a**) Schematic of processing the heat equation: The heat source parameters and Laplacian matrix data undergo high-speed loading onto the NTPC. Following computation, the optical outputs are converted to electrical signals by a photodetector (PD), captured via an oscilloscope, and processed computationally. (**b**) Laplacian operator coefficient matrix with a grid size of 36×36. (**c**) Signal waveform generated from the dot product of the heat source matrix and the discrete Laplacian operator (computation: gray vs experiment: blue). (**d**) Simulated and (**e**) experimental initial thermal fields. (**f**) Simulated and (**g**) experimental thermal field distributions at 0.75s, 2s,

and 3.5s. (**h**) Dynamic changes in the thermal field at the grid point (5,5). (**i**) Time-dependent solution precision from 0 to 5 s, achieving over 94.5% accuracy.

**Reconfigurable NTPC for high bandwidth density optical I/O.** Beyond enhancing computational density in optical computing, the ultra-compact topological modulator is particularly suited for short-reach optical interconnects in future disaggregated data centers, where extreme compactness and high speed are critical[40]. While micro-ring modulators currently serve as core components, our topological modulator achieves comparable bandwidth and speed while being two orders of magnitude smaller.

Leveraging the non-volatility of PZT, we demonstrate near-zero-power dynamic reconfiguration of the NTPC. By precisely engineering 16 back-end TPC modulators within the NTPC (Fig. 6a), DWDM is realized with 100 GHz channel spacing (Figs. 6b-e). High-speed testing reveals well-defined eye diagrams for on-off keying (OOK) signals across all 16 modulators operating at 50 Gbps (Fig. 6f), achieving an aggregate data throughput of 0.8 Tbps. In contrast to microring-based modulators from Ayar Labs[41] and Intel[42], which demonstrate bandwidth densities of 0.36 Tbps/mm and 0.46 Tbps/mm, respectively, our NTPC harnesses the miniaturization capabilities of topological photonics to achieve a bandwidth density of 3.56 Tbps/mm within a 0.225-mm-long shoreline footprint. This metric—commonly used to evaluate optical I/O capacity—highlights how linear scaling directly influences channel scalability, underscoring the superior integration density achieved by our platform. Detailed experimental validations are provided in the Supplementary Section [12] and Supplementary Section [13].

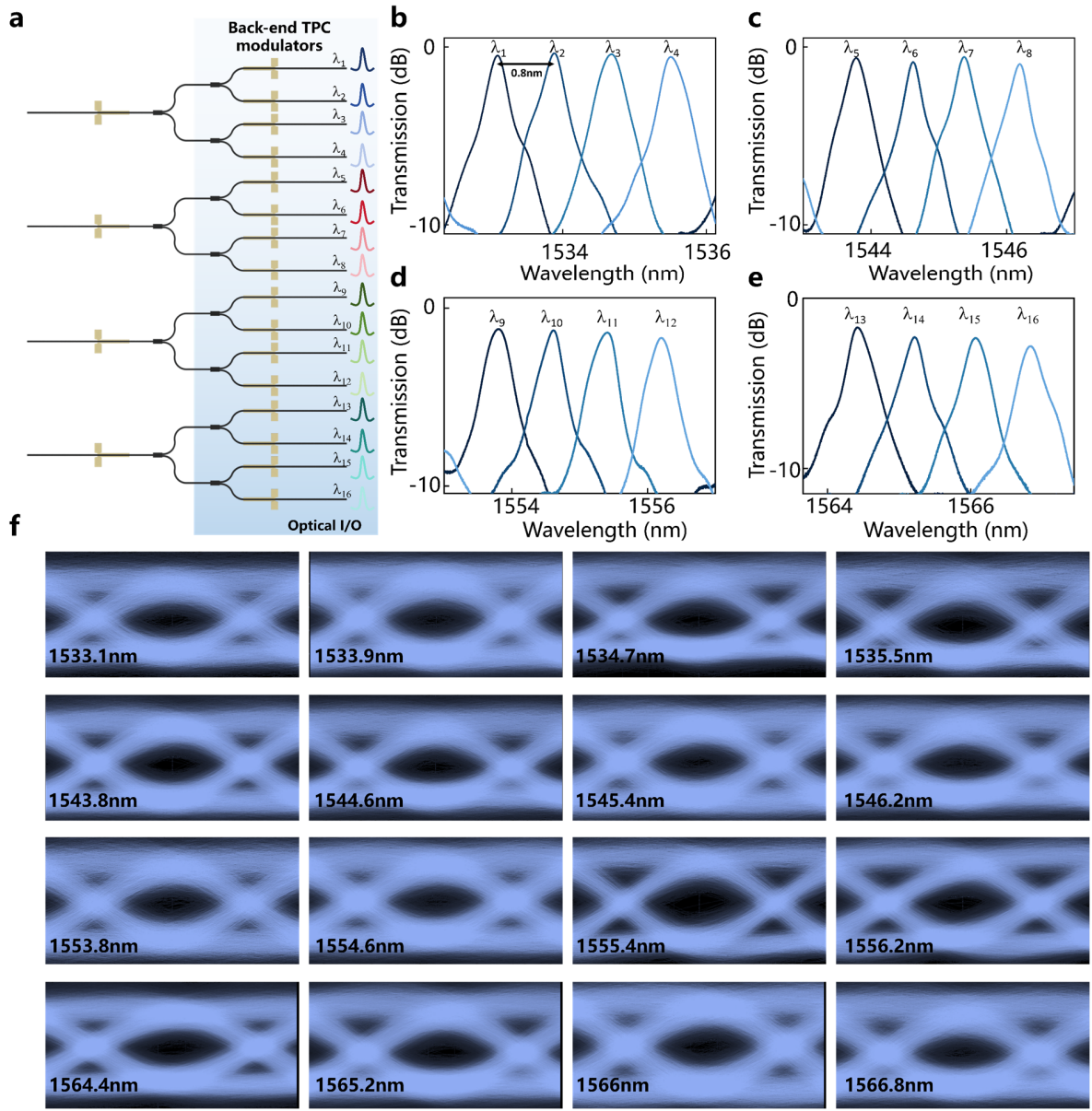

**Fig. 6. Reconfigurable NTPC for DWDM systems.** (**a**) Schematic of the DWDM system reconfigured via NTPC architecture. (**b-e**) Non-volatile transmission spectrum reconfiguration with 100 GHz (0.8 nm) uniform channel spacing. (**f**) Measured eye diagrams of OOK signals for all 16 TPC modulators operating at 50 Gbps.

**Table 1. Performance comparison of state-of-the-art reconfigurable optical computing chips with different platforms and architectures.**

| Platform | Architecture | Rate (GBaud) | Compute efficiency (fJ/OPS) | On-chip efficiency (fJ/OPS) | Compute density (TOPS/mm$^2$) | Dimension of vector in product | Precision | Accuracy on MNIST |
|---|---|---|---|---|---|---|---|---|
| TFLN | Dual MZIs[28] | 60 | 213 | 56.37 | 0.012[a] | 1.3×10$^5$ | 6-bit | 92% |
| | Cascaded MZI[43] | 20 | 1250 | 66.6 | 0.008[a] | 16 | / | 88.5% |
| | MZI+ MRR[44] | 18.35 | 6950 | 2550 | 0.031 | / | 5-bit | 88% |
| Si | Comb+ MRR[45] | 17 | 6.58×10$^5$ | 3.72×10$^4$ | 1.04[a] | 4 | 9-bit[b] | 96% |
| | Cascaded MZI[31] | / | / | 30[c] | 1.12[a] | 16 | 5-bit | 76.7% (4 categories, vowel recognition) |
| SiN | Comb+ PCM[5] | 2 | 500 | / | 1.2[a] | 9 | 7-bit | 95.3% |
| PZT-SiN (This work) | Dual TPCs | 60 | 265 | 6.25 | 266 | 4.2×10$^6$ | 5-bit | MNIST classification/ 95.64% and 2D time-varying PDEs/94.5% |

a. Calculated from supplementary materials; b. For consistency, the data is recalculated using the standard deviation provided in the supplementary material; c. These data can be obtained based on existing state-of-the-art equipment.

## Discussion

In conclusion, we have designed and realized a programmable topological photonic chip that synergistically incorporates non-volatile tunability, ultrafast electro-optic modulation, and strong optical confinement within a compact architecture. By leveraging the large Pockels coefficient and non-volatile characteristics of ferroelectric PZT thin films, we demonstrate sub-nanosecond reconfiguration of topological edge states with near-zero static power consumption. The chip delivers exceptional performance in two key domains: optical computing and optical interconnects. It supports parallel photonic in-memory computing with a computational density of 266 TOPS/mm²—surpassing conventional reconfigurable computing architectures by two to four orders of magnitude. Simultaneously, it serves as a high-bandwidth DWDM interface with an I/O shoreline density exceeding 3.56 Tbps/mm, confirming superior scalability and versatility.

Our results establish the high-speed, non-volatile ferroelectric platform outperforming existing solutions in speed, integration density, and energy efficiency. The chip's flexibility in fan-in/fan-out scaling and rapid weight updates positions it as a compelling candidate for next-generation photonic computing (Table 1). Looking ahead, integrating optical microcombs on the PZT platform could dramatically expand parallel wavelength channels, enhancing photonic convolution capabilities for complex AI tasks. Co-design with high-speed electronic interfaces (e.g., analog-to-digital convertors / digital-to-analog convertors) promises further reductions in system power and latency. Scaling to larger modulator arrays and optimizing topology-specific algorithms will be critical to harness NTPC's full potential for real-time scientific computing, edge intelligence, and beyond.

## Methods

**Device fabrication**

The fabrication process of the NTPC is as follows: First, a 300 nm-thick PZT ferroelectric thin film is deposited on a substrate consisting of a 2 μm-thick $SiO_2$ insulating layer and a 525 μm-thick Si substrate using the chemical solution deposition (CSD) method. Subsequently, a 400 nm-thick SiN dielectric layer is deposited via plasma-enhanced chemical vapor deposition (PECVD, Oxford). For patterning, the silicon nitride waveguide structure is precisely defined using a Vistec EBPG 5200+ electron-beam lithography (EBL) system with AR-P6200.09 resist as the mask, followed by pattern transfer into the SiN layer via inductively coupled plasma (ICP) dry etching. Finally, electrodes and bonding pads are fabricated by electron-beam evaporation, where a 10 nm-thick Ti adhesion layer is first deposited, followed by a 300 nm-thick Au layer, with the final metal structures formed through a lift-off process.

**Author contributions**

Conceptualization: W Zhou, X Wang, X Zhang, Y Zhang. Methodology: W Zhou, X Wang, X Zhang, Y Zhang. Investigation: W Zhou, X Wang, X Zhang, Y Chen, M Sun, J Li, X Ni. Visualization: W Zhou, X Wang, X Zhang, Y Zhu, Q Han, J Wang, C Yang, B Li. Funding acquisition: Y Zhang, Y Su. Project administration: Y Zhang, Y Su. Supervision: Y Zhang, Y Su, F Qiu. Writing - original draft: W Zhou, X Wang, X Zhang, Y Zhang, Y Su. Writing - review & editing: W Zhou, X Wang, X Zhang, Y Zhang, Y Su.

**Acknowledgements**

This work was supported by the National Natural Science Foundation of China (NSFC) under Grant 62335014 and the Shanghai Municipal Science and Technology Commission Project under Grant 25JD1402000. We thank the Center for Advanced Electronic Materials and Devices (AEMD) of Shanghai Jiao Tong University (SJTU) for their support in device fabrication.

**Competing interests**

The authors declare no competing interests.

**Data availability**

The data supporting the plots in this study and other findings of this study are available from the corresponding author upon reasonable request.